\documentclass[5p,sort&compress,number]{elsarticle} 
\usepackage[english]{babel}
\usepackage{epsfig}
\usepackage{array}
\usepackage{amsmath,amssymb}
\usepackage{verbatim} 
\usepackage{lineno}   
\usepackage{graphicx}  
\usepackage[hyperindex,breaklinks]{hyperref}

\begin{document}

\begin{frontmatter}
\title{Anisotropy of TeV and PeV cosmic rays with IceCube and IceTop}
 
\author[WI]{M.\ Santander}
\author{for the IceCube Collaboration\corref{url}}
\address[WI]{University of Wisconsin-Madison, Madison, WI 53703, USA}
\cortext[url]{\href{http://www.icecube.wisc.edu}{http://www.icecube.wisc.edu}} 
 
 \begin{abstract}
 The interaction of high energy cosmic rays with the Earth's atmosphere produces extensive
 air showers of secondary particles with a large muon component. By exploiting the sensitivity
 of neutrino telescopes to high energy muons, it is possible to use these detectors for precision
 cosmic ray studies. The high rate of cosmic-ray muon events provides a high-statistics data 
 sample that can be used to look for anisotropy in the arrival directions of the parent particles at the 
 per-mille level.
 This paper reports on the observation of anisotropy in the cosmic ray data collected with the
 IceCube neutrino telescope in the 20-400 TeV energy range at multiple angular scales. New 
 data from the IceTop air shower array, located on the ice surface above IceCube, shows an 
 anisotropy that is consistent with the high-energy IceCube results. The sensitivity of 
 IceTop to all the components of the extensive air shower will allow us to explore in more 
 detail the characteristics of the primary cosmic rays associated with the observed anisotropy.
 \end{abstract}
 
 \begin{keyword}
 galactic \sep cosmic ray \sep anisotropy 
 \end{keyword}

\end{frontmatter}

\section{Introduction}

It has been almost 100 years since the discovery of cosmic rays by Victor Hess. The origin of these
energetic particles, however, still remains an enduring problem in astrophysics. Based on indirect 
evidence, it is believed that cosmic rays (CRs) with energies up to a few PeV are accelerated in 
supernova remnants distributed accross our galaxy. 
A direct test of this hypothesis is a challenge since the arrival directions of cosmic rays at Earth
do not point back to their sources due to the scrambling action of the galactic magnetic field (GMF)
during propagation. For this reason, the search for the sources of cosmic rays is usually performed either
at the highest energies where the influence of the GMF is small, or by making use of a neutral messenger 
particle as the neutrino.

Even if a direct detection of cosmic ray sources is not feasible at or below PeV energies, their 
discrete spatial distribution should create an observable dipolar anisotropy of per-mille strength
\citep{Erlykin:2006ri}\citep{Ptuskin:2006aa}\citep{Blasi:2011aa}. The energy dependence of the phase and amplitude 
of this kind of anisotropy would be dominated by details in the propagation process such as the geometry of the 
galaxy, the energy dependence of the CR diffusion coefficient, and the age and injection spectrum of 
the sources. Other factors, such as turbulent propagation  in the GMF \citep{Giacinti:2011aa}, 
heliospheric effects \citep{Lazarian:2010sq}, and special magnetic field configurations 
\citep{Malkov:2010yq}\citep{Drury:2008ns}, could give rise to anisotropy at smaller angular scales.
A different process, known as the Compton-Getting effect \citep{Compton:1935}, could also create 
a dipole anisotropy due to the relative motion of the solar system with respect to the cosmic ray 
plasma.

Several experiments in the northern hemisphere have reported on the observation of anisotropy at 
TeV energies \citep{Abdo:2008kr}\citep{Aglietta:2009mu}\citep{Amenomori:2006bx}\citep{Guillian:2007}\citep{Vernetto:2009xm}. 
Two features dominate the northern sky in cosmic rays: a dipole-like large 
scale structure with an amplitude of $\sim 10^{-3}$, and a small scale anisotropy with significant 
structure at angular sizes between $10^{\circ}$ and $30^{\circ}$. The observed dipole anisotropy is inconsistent,
both in amplitude and phase, with the Compton-Getting prediction.

IceCube is sensitive to muons from cosmic rays with TeV energies, and the data collected with this detector has been used 
to provide the first look at the CR anisotropy in the southern sky. The large set of cosmic ray events from IceCube, together 
with the air shower data from the IceTop detector, provide us with important tools to study the 
anisotropy of cosmic rays in the TeV and PeV range.

The most recent results on CR anisotropy obtained with the IceCube and IceTop detectors are summarized
in this paper.
 
\section{The IceCube and IceTop detectors}

IceCube is a km$^3$ neutrino telescope designed to search for astrophysical sources of high energy 
neutrinos. The basic building block of IceCube is the Digital Optical Module (DOM), a glass pressure 
sphere that contains a 10" Hamamatsu PMT \cite{Abbasi:2010vc}, together with electronic boards for signal digitization \cite{Abbasi:2008ym}, HV 
supply, and calibration LEDs. Between 2004 and 2010, 5160 DOMs were deployed in the South Pole ice at
depths between 1450 and 2450 m to detect the Cherenkov radiation emitted by charged particles produced in the 
interaction of neutrinos with nucleons.

These DOMs are attached to 86 vertical strings that provide mechanical support, electrical power, and a data 
connection to the surface. The vertical spacing between most consecutive DOMs in each string is about 17 
m, while the horizontal spacing between most neighboring strings is approximately 125 m.  

A dedicated cosmic ray air shower array called IceTop is located on the ice surface above IceCube.
The array consists of 81 stations, with two light-tight ice Cherenkov tanks per station. Each tank 
is 1.8 m in diameter, 1.3 m in height, and is instrumented with two DOMs that are operated at 
different PMT gains to increase the dynamic range of the detector.

IceCube and IceTop were operated in partial configurations from the beginning of their 
construction until the completion of both detectors in December of 2010.
IceCube was operated in a 59-strings (IC59) configuration between 
May 2009 and May 2010, with IceTop operating with 59 stations (IT59) during the same time period.

The IC59 dataset consists of those events where at least 8 IceCube DOMs detected photon hits within 
a 5 $\mu$s window. The average rate for this trigger condition was 1.7 kHz in IC59. During a total live time 
of 334.5 days, $3.4 \times~10^{10}$ events were collected, almost all
of them produced by down-going muons from cosmic rays. For this analysis, a fast muon track reconstruction
was performed online at the South Pole. The result of the fit, together with the number of triggered
DOMs and the time of the event are stored and transferred over a satellite link using a compressed data format.
The median energy of primary cosmic rays in this dataset is 20 TeV and was determined through Monte Carlo 
simulations assuming a mixed CR composition dictated by the polygonato model \citep{Hoerandel:2002yg}. 
In this model, the energy spectrum for each chemical element is given by a broken power-law with a smooth 
transition, where the location of the spectral break is rigidity-dependent. Due to this dependence, heavier elements
dominate the all-particle CR spectrum at energies above a few PeV.

The median angular resolution of the muon track reconstruction is 3$^{\circ}$. Due to the degradation of 
the resolution with increasing zenith angle, only events with $\theta < 65^{\circ}$ were used in the analysis, 
which reduced the final dataset size to 3.2 $\times 10^{10}$ events. 

In IceTop, the high-gain DOMs in the two tanks that form a station are run in local coincidence mode
and the readout is enabled if they record hits within $\pm 1~\mu$s of each other. The IceTop trigger 
condition is satisfied if at least 6 DOMs recorded locally-coincident hits within a time window of 5
$\mu$s, which implies that at least 2 stations have participated in the event. 

The anisotropy analysis used events in which at least 3 IceTop stations had triggered. Due 
to bandwidth limitations, events triggering less than 8 stations were prescaled by a factor of 8 while 
events with at least 8 stations were not prescaled. The event directional reconstruction was 
performed doing a $\chi^2$ fit to the trigger times of each station using a planar approximation for 
the shape of the shower front. Simulations show that the median resolution of this reconstruction algorithm 
is 2$^{\circ}$. 

Preliminary results from Monte Carlo studies using a mixed composition of H, He, and Fe from the polygonato model
indicate that the median primary CR energy of the IceTop dataset is 640 TeV, with 68\% of the
event between 200 TeV and 2400 TeV. Only events with $\theta < 60^{\circ}$ were selected for the analysis, with 
$1.2 \times~10^8$ events passing the cut.

\section{Analysis and results}

An anisotropy in the arrival direction of TeV cosmic rays was observed for the first time in IceCube 
using data from the 22-string configuration (IC22) that operated between June 2007 and March 2008 
and was reported in Ref. \cite{Abbasi:2010mf}. In this analysis, the exposure-corrected right ascension distribution 
of cosmic ray events was fitted with a harmonic function of the form $\sum_i A_i \cos(i(\alpha -\phi_i)) + 
B$, where $A_i$ and $\phi_i$ are the amplitude and phase of the $i^{\mathrm{th}}$ term in the sum, 
$\alpha$ is the right ascension, and $B$ is a constant. This sum was performed over the first two terms 
in harmonic space ($n=1,2$) since they provide with an adequate description of the shape of the anisotropy.
The fit parameters obtained in this analysis are $A_1 = (6.4 \pm 0.2_{\mathrm{(sta)}} \pm 0.8_{\mathrm{(sys)}}) 
\times 10^{-4}$, $\phi_1 = 66.4^{\circ} \pm 2.6^{\circ}_{\mathrm{(sta)}} \pm 3.8^{\circ}_{\mathrm{(sys)}} $,
$A_2 = (2.1 \pm 0.3_{\mathrm{(sta)}} \pm 0.5_{\mathrm{(sys)}}) \times 10^{-4}$, $\phi_2 = -65.6^{\circ} \pm 
4.0^{\circ}_{\mathrm{(sta)}} \pm 7.5^{\circ}_{\mathrm{(sys)}} $ with $\chi^2/\mathrm{dof} = 22/19$, and show
a good agreement with the phase and amplitude of the anisotropy observed in the northern sky.

A later analysis \cite{Abbasi:2011aa} using IC59 data revealed that besides the large-scale structure 
(i.e. dipole and quadrupole modes) observed in the IC22 analysis there are also statistically significant 
structures with typical sizes between 10$^{\circ}$ and 20$^{\circ}$.
In this analysis, the search for anisotropy is conducted by searching for deviations of the sky map of 
reconstructed cosmic ray arrival directions in equatorial coordinates from a reference isotropic sky map 
obtained from data using the time-scrambling method described in Ref. \citep{Alexandreas:1993}. 
The time scrambling period used in the analysis is 24 hours, which makes it sensitive to 
all angular scales in the celestial sphere. During the time scrambling procedure, events were resampled 
20 times to reduce statistical fluctuations in the reference sky map.
  
The sky maps were constructed using the HEALPix\footnote{\href{http://healpix.jpl.nasa.gov}{http://healpix.jpl.nasa.gov}}
library \citep{Gorski:2004by} that provides an equal area pixelization of the sphere. The chosen HEALPix 
resolution divides the sphere into 49152 pixels, with an average distance between pixel centers of 
approximately 1$^{\circ}$. Using the reference and data maps, a relative intensity map can be calculated 
using the expression $\delta I_i = (N_i - \langle N \rangle_i) / \langle N \rangle_i$, where $N_i$ and 
$\langle N \rangle_i$ are respectively the number of observed events and the number of reference events
for the isotropic expectation in the $i^{\mathrm{th}}$ pixel obtained with the time scrambling technique.

 \begin{figure}[h]
  \centering
  \includegraphics[width=.45\textwidth]{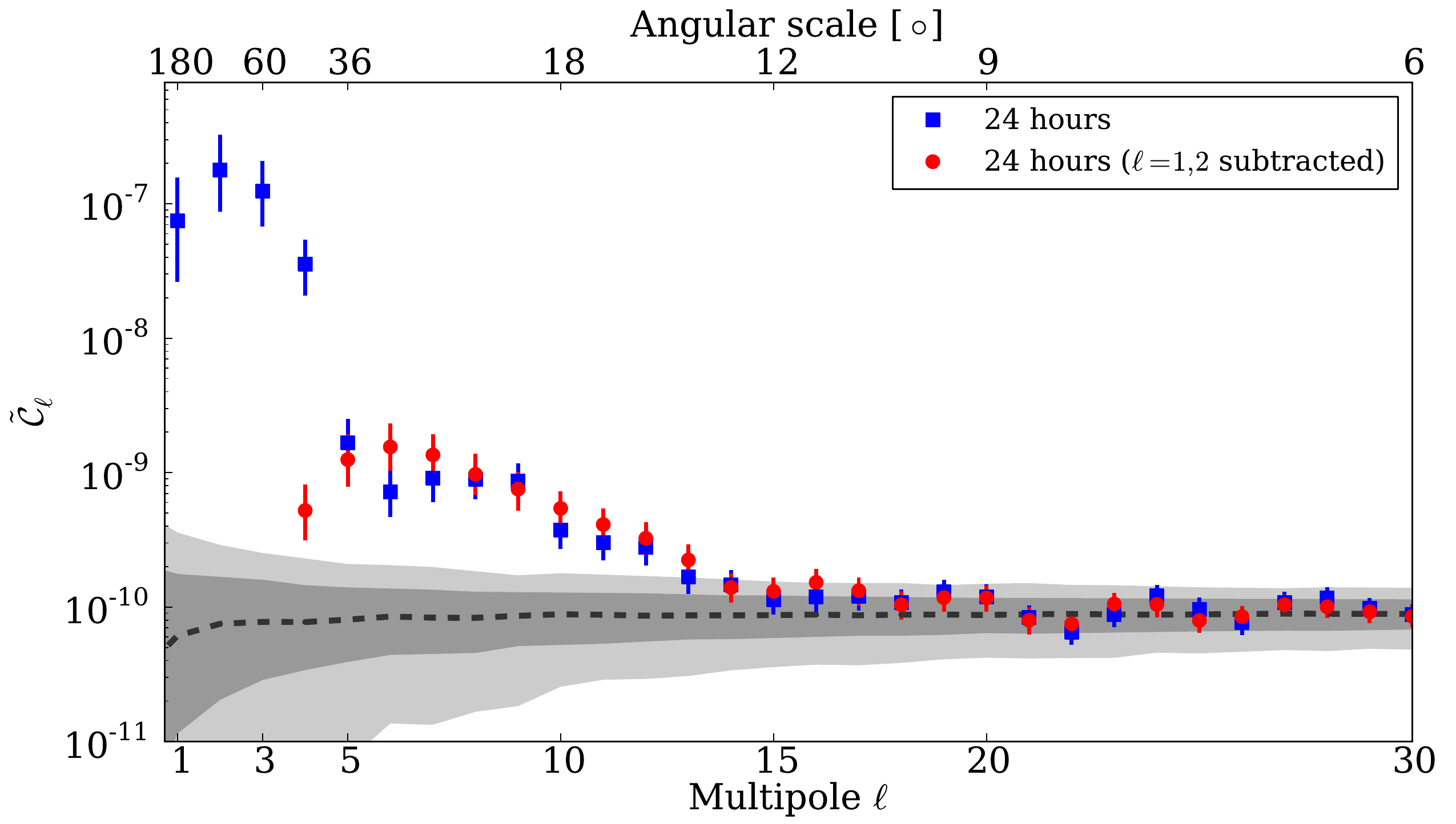}
  \caption{Angular power spectrum of the IC59 relative intensity skymap. The power spectrum before 
  (\emph{blue}) and   after (\emph{red}) the subtraction of the dominant dipole ($\ell = 1$) and quadrupole 
  ($\ell = 2$) terms are shown for a time scrambling period of 24 hours. The median value for the isotropic 
  expectation is shown as a dashed black line, while the 1$\sigma$ and 2$\sigma$ bands are shown in gray. 
  The power spectrum  shows significant departures from isotropy between $\ell \sim  4$ and $\ell 
  \sim 12$ even after the subtraction.
  of the low-order terms.}
  \label{f:powspectrum}
\end{figure} 

\begin{figure}[h]
  \centering
  \includegraphics[width=.45\textwidth]{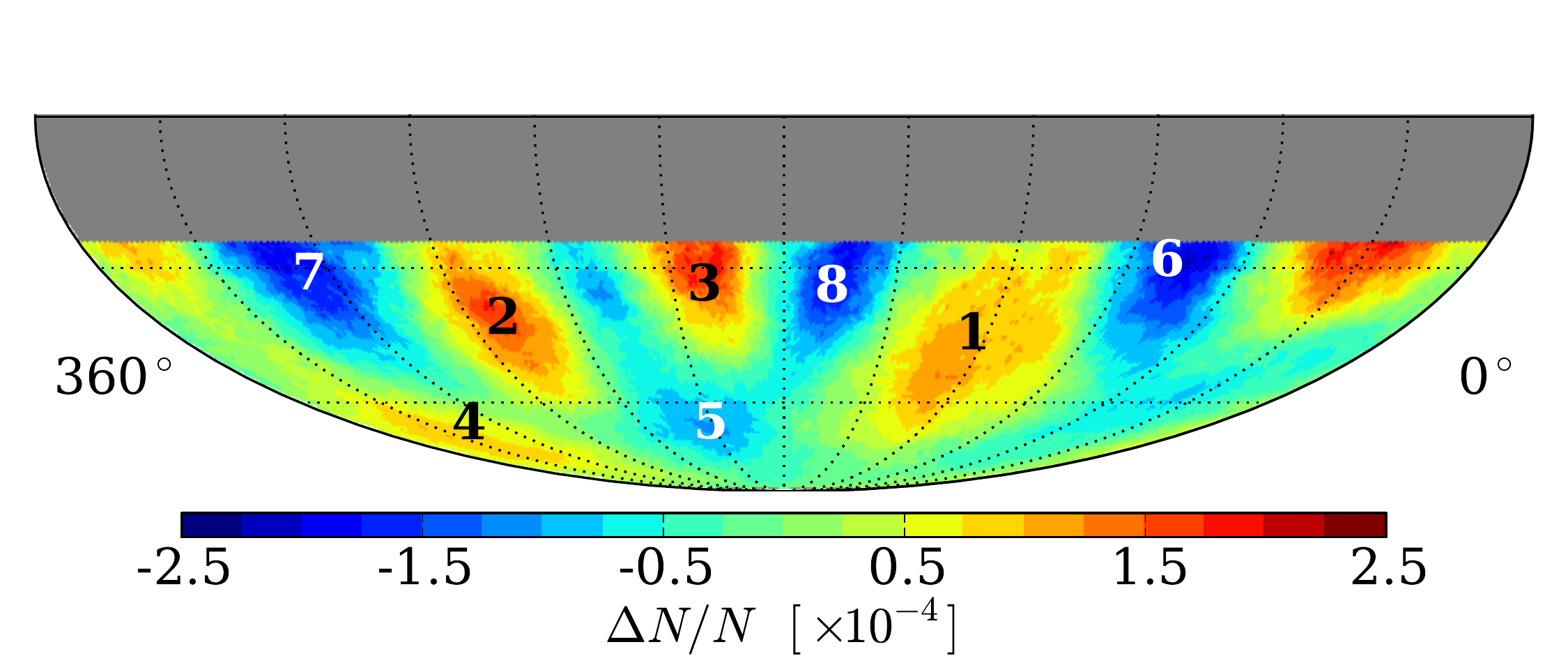}
  \caption{Relative intensity sky map of the residual small scale structure after the subtraction of a dipole and quadrupole terms for a
  smoothing radius of $15^{\circ}$. The labels correspond to the locations of all regions that showed an statistical significance larger than $5\sigma$. 
  See Ref.~\cite{Abbasi:2011aa} for a full list of coordinates.  }  
  \label{f:ic59small}
\end{figure}

The angular power spectrum of the relative intensity map can be used to estimate the strength 
of the anisotropy at different angular scales in our data. The IC59 power spectrum is shown in 
Fig.~\ref{f:powspectrum} and was obtained using the \texttt{PolSpice} software package\footnote{\href{http://www2.iap.fr/users/hivon/software/PolSpice/}
{http://www2.iap.fr/users/hivon/software/PolSpice/}} that corrects for systematic effects introduced by the 
limited sky coverage of our data \cite{Chon:2003gx}\cite{Szapudi:2000xj}. It can be seen that besides
the already mentioned dipole ($\ell =1$) and quadrupole ($\ell = 2$) there is a significant departure from 
isotropy at higher multipole moments between $\ell \sim 6$ and $\ell \sim 12$, which corresponds to 
structures that have angular sizes roughly between 15$^{\circ}$ and 30$^{\circ}$ in the sky. 
In order to reveal this smaller structure, the dipole ($\ell = 1$) and quadrupole ($\ell = 2$) terms of the
spherical harmonics functions were fit and subtracted from the IC59 relative intensity map.
The residual maps were smoothed to search for the small scale anisotropy. The smoothing procedure sums all 
events in a pixel to the events from pixels inside a certain angular distance. This produces a sky map of correlated 
pixels with an improved sensitivity to structures with angular sizes similar to the smoothing radius. A scan was then
performed over smoothing radii between 3$^{\circ}$ and 30$^{\circ}$ in $1^{\circ}$ steps to find the optimal angular 
scale for the small scale structure. The optimal scale corresponds to the one that maximizes the statistical significance
of the observation calculated according to Ref.~\cite{LiMa:1983}, after taking into account trials due to the search 
over many pixels and smoothing radii. For the IC59 data, eight regions were identified where the absolute value 
of the statistical significance was higher than $5\sigma$ before accounting for trials. These regions can be seen 
in Fig.~\ref{f:ic59small} for smoothing radii of $15^{\circ}$. The most significant excess is region 1, with an optimal 
scale of $22^{\circ}$ at which it reaches a significance of $5.3\sigma$ after trials and has an amplitude of $\sim 10^{-4}$. 
A full list of the statistical significance for all regions can be found in Ref. \cite{Abbasi:2011aa}.

\begin{figure}[h]
 \begin{center}
 \includegraphics[width=.45\textwidth]{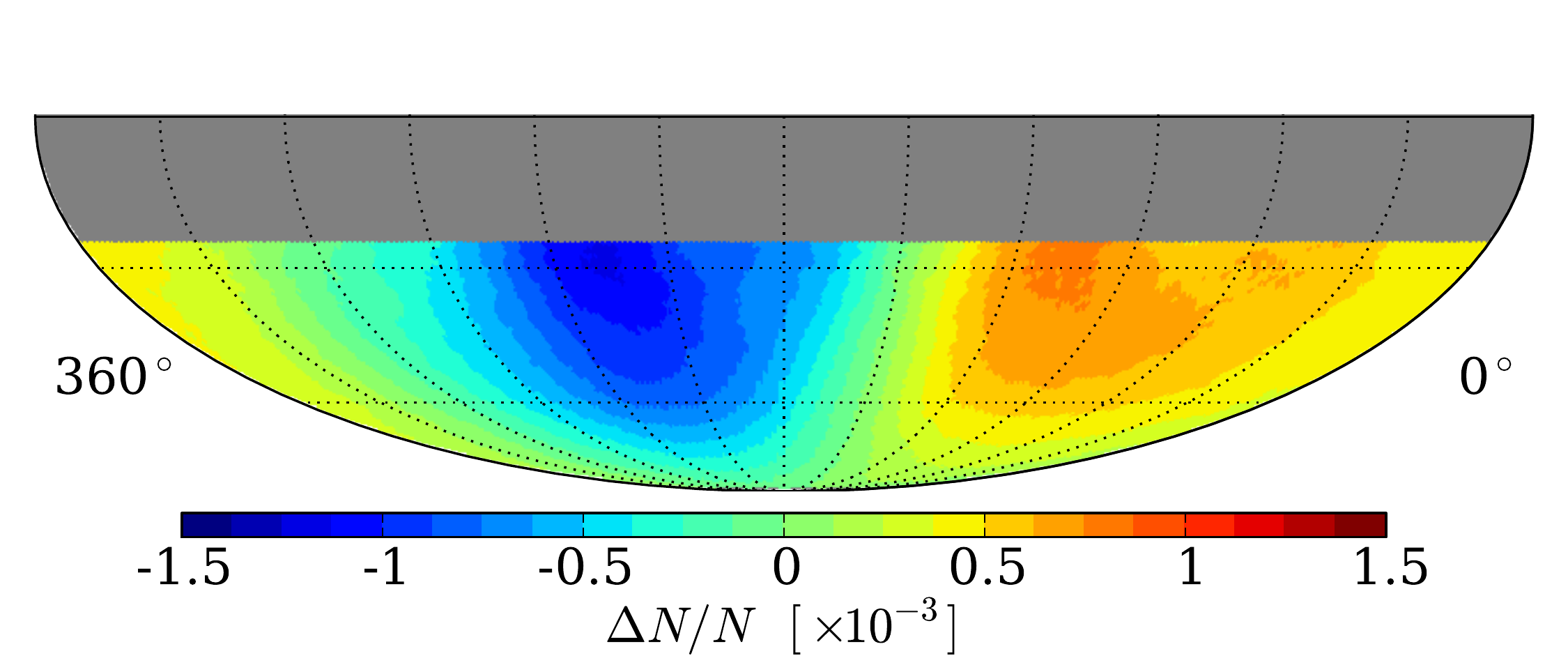}
 \includegraphics[width=.45\textwidth]{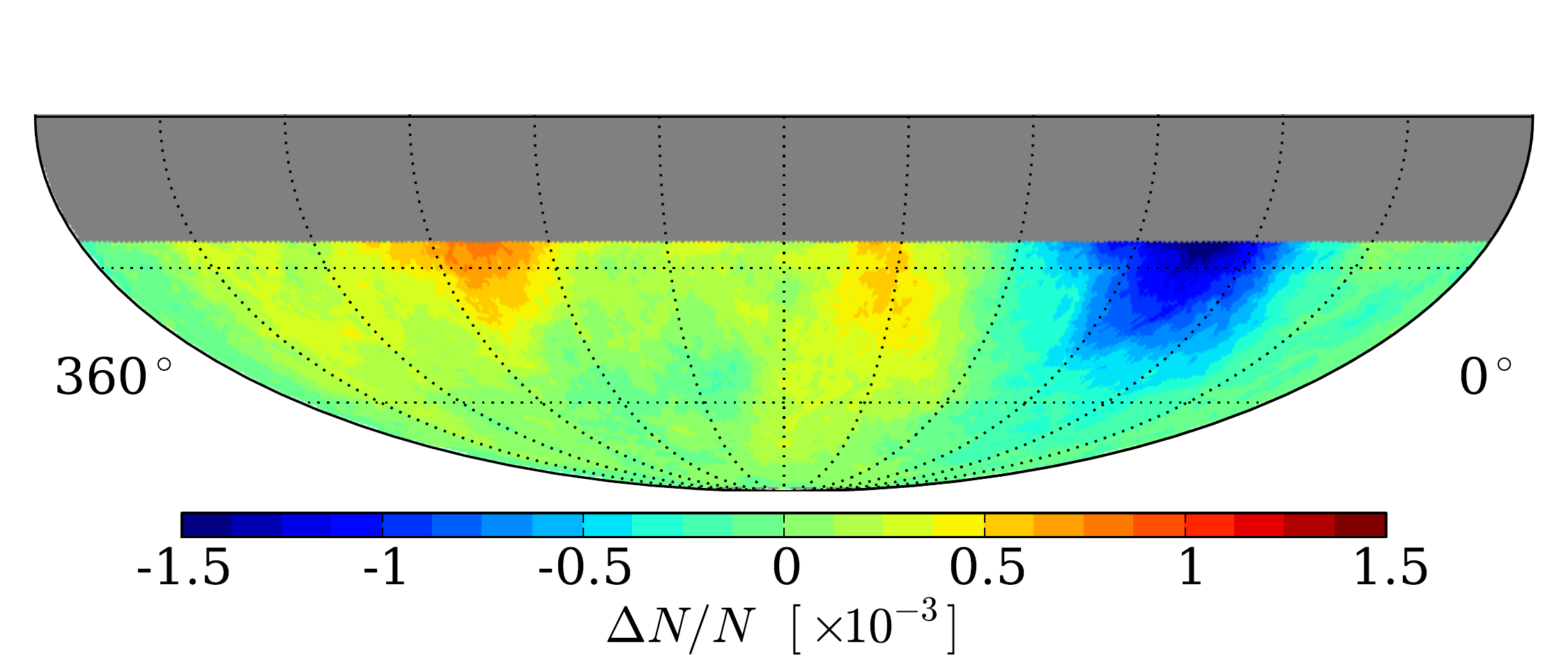}
 \end{center}
 \caption{Equatorial sky maps for relative intensity in two different energy bands using the IC59 dataset:
 20 TeV (\emph{above}), and 400 TeV (\emph{below}) for a smoothing radius of $20^{\circ}$.}
 \label{f:edep}
\end{figure}

A study of the energy dependence of the anisotropy was performed using IC59 data and was reported in Ref.
\cite{Abbasi:2011ab}. A cut was implemented to create two distinct subsamples with different median 
energies: 20 TeV, and 400 TeV. The cut variables used in the selection were the reconstructed 
zenith angle of the event, and the number of triggered DOMs (both increase as a function of primary CR energy).
 After cuts, the 20 TeV dataset contained $17.9 \times 10^9$ events, while the 400 TeV dataset consists 
 of $0.5 \times 10^9$ events (with 68\% of the events  between 100 TeV and 1300 TeV).
The anisotropy results were obtained in two ways: through a harmonic fit to the right ascension distribution
of events as in the case of IC22, and through the search for the optimal angular scale after a reference level 
estimation performed with the time scrambling technique as in  Ref. \cite{Abbasi:2011aa}. 
Both methods consistently observed the presence of the already known dipole and quadrupole structure
in the 20 TeV dataset, while at 400 TeV the anisotropy pattern changes both in phase and amplitude. 
Only one structure in the 400 TeV sky map has a post-trial significance larger than $5\sigma$: a $6.3\sigma$ 
deficit located at ($\alpha = 73.1^{\circ}, \delta=-25.3^{\circ}$) with an optimal smoothing 
of $21^{\circ}$ and an average amplitude of approximately $7 \times 10^{-4}$ in relative intensity. The deficit
is also visible as region 6 in the small scale map shown in Fig.~\ref{f:ic59small} before energy cuts are applied. 
This observation represents the first detection of anisotropy in the southern sky at these energies.
The relative intensity sky maps for the 20 TeV and 400 TeV energy bands are shown in Fig.~\ref{f:edep}.

 A preliminary analysis of the IceTop IT59 dataset reveals a deficit located in the same region as the one 
 observed at 400 TeV with IceCube.  For a smoothing angle of $20^{\circ}$ the pre-trial significance is 
 $6.2 \sigma$ and the amplitude is about $2 \times 10^{-3}$, larger than the one observed in 
 IceCube. A possible cause of this discrepancy is the difference in energy range associated with the two 
 data sets. It is also possible that the CR chemical composition may be contributing to this mismatch. 
 This is due to the fact that while IceTop is sensitive to all components of the CR air shower, IceCube 
 is only capable of detecting the muon component, and this could create a detection bias towards a 
 particular composition. Further studies of the energy and composition dependence of the 
 anisotropy are needed to perform a direct comparison between both results.
  
\begin{figure}[h]
 \begin{center}
 \includegraphics[width=.5\textwidth]{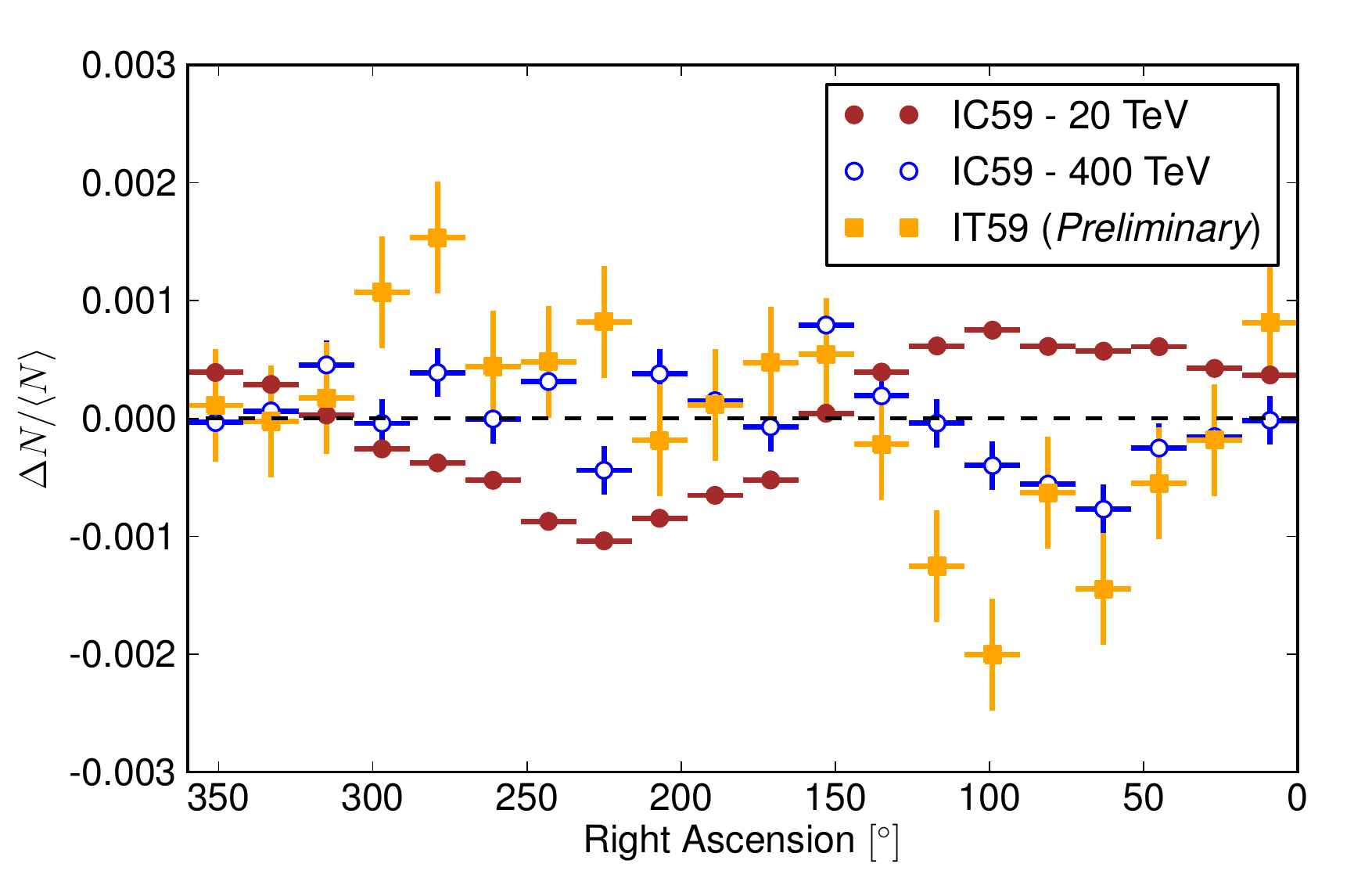}
 \end{center}
 \caption{Relative intensity as a function of right ascension for the IC59 20 TeV (\emph{brown}), and IC59 400 TeV (\emph{blue}) 
 compared to a preliminary result for the IceTop IT59 (\emph{orange}) datasets. For clarity, only statistical error bars are shown.}
 \label{f:ra1d}
\end{figure}

 A comparison of the right ascension projection of the relative intensities 
 observed in IceCube for both the 20 TeV and 400 TeV energy bands, and IceTop is shown in Fig.~\ref{f:ra1d}. 
 Only events in the $-75^{\circ} \le \delta \le -30^{\circ}$  declination range were used in this plot.
 
\begin{figure}[h]
 \begin{center}
 \includegraphics[width=.45\textwidth]{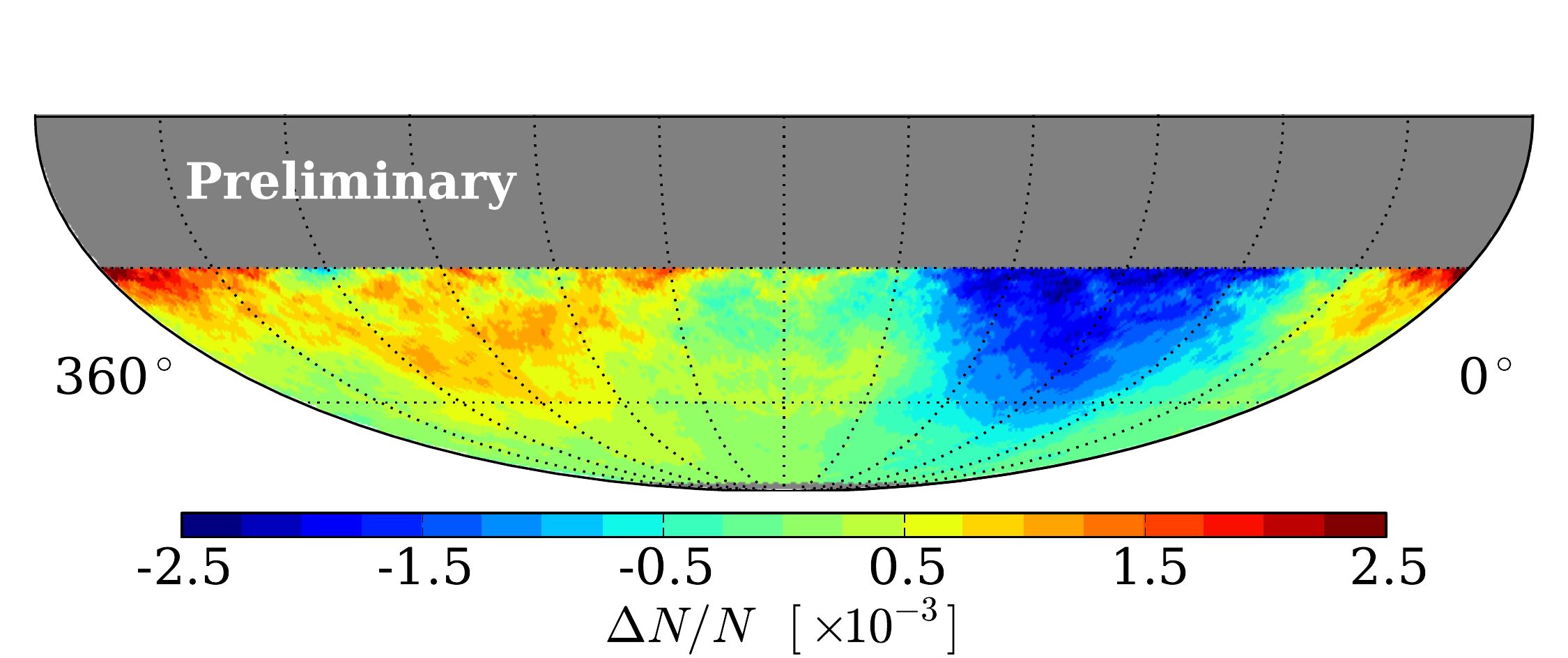}
 \end{center}
 \caption{Relative intensity sky map in equatorial coordinates for the IceTop IT59 dataset with a smoothing radius
 of $20^{\circ}$ (\emph{preliminary}).}
 \label{f:icetop}
\end{figure}

\section{Summary}

Data taken between 2007 and 2010 with the IceCube neutrino telescope and the IceTop air shower array has 
been used to probe the anisotropy of TeV and PeV cosmic rays down to amplitudes of $10^{-4}$. The 
anisotropy at an energy of 20 TeV is consistent with that observed by other experiments in the northern hemisphere,
 and is dominated by a large scale component (dipole and quadrupole) with a strength of $\sim 10^{-3}$. A subdominant, 
 but statistically  significant, structure at 20 TeV is characterized by small excess and deficit regions with angular sizes 
 between  $10^{\circ}$ and $25^{\circ}$ and strengths of the order of $10^{-4}$.

At energies of about 400 TeV,  IceCube observed a strong deficit with a relative intensity of about $10^{-3}$ and 
a size of approximately $20^{\circ}$. A preliminary analysis of IceTop data shows a deficit in the same region, but with
an amplitude that doubles the one observed in IceCube. 

Future studies will expand the energy reach of the anisotropy analysis, and provide a handle on the evolution of the
anisotropy as a function of energy and angular scale.


\end{document}